\documentclass[10pt]{article} 
\usepackage[preprint]{rlc}

\usepackage{amssymb}            
\usepackage{mathtools}          
\usepackage{mathrsfs}           
\mathtoolsset{showonlyrefs}     
\usepackage{graphicx}           
\usepackage{subcaption}         
\usepackage[space]{grffile}     
\usepackage{url}                

\usepackage{wrapfig}
\usepackage{hyperref}
\usepackage{booktabs}
\usepackage{xspace}
\usepackage{pifont}
\usepackage{xcolor}
\usepackage{multirow}
\usepackage{colortbl}
\usepackage{tabularx}
\usepackage{enumitem}
\usepackage{adjustbox}
\usepackage{ragged2e}   
\usepackage{makecell} 
\usepackage{array}       
\usepackage{tcolorbox}
\usepackage{listings}
\usepackage{tikz}

\usepackage{arydshln}

\newcommand{\gradientColorAgent}{%
    \textcolor{red!10!orange}{C}%
    \textcolor{red!50!orange}{o}%
    \textcolor{magenta!70!red}{l}%
    \textcolor{purple!80!magenta}{o}%
    \textcolor{purple!40!violet!70!magenta}{r}%
    \textcolor{purple!30!violet!70!magenta!80!blue}{A}%
    \textcolor{purple!60!magenta!50!blue}{g}%
    \textcolor{purple!60!magenta!35!blue}{e}%
    \textcolor{purple!60!magenta!30!blue!80!cyan}{n}%
    \textcolor{purple!60!magenta!30!blue!60!cyan}{t}%
}

\title{\gradientColorAgent: Building A Robust, Personalized, and Interactive OS Agent}

\author{\\
\name{Ning Li}$^{1}$\thanks{Equal contribution.}~\thanks{Work done during an internship at OPPO.}~,
\name{Qiqiang Lin}$^{2}$\footnotemark[1]~,
\name{Zheng Wu}$^{1}$\footnotemark[2]~,
\name{Xiaoyun Mo}$^{2}$,
\name{Weiming Zhang}$^{1}$\footnotemark[2]~, 
\name{Yin Zhao}$^{2}$, \\
\name{Xiangmou Qu}$^{2}$,
\name{Jiamu Zhou}$^{2}$,
\name{Jun Wang}$^{2}$,
\name{Congmin Zheng}$^{1}$\footnotemark[2]~, 
\name{Yuanyi Song}$^{1}$\footnotemark[2]~, \\
\name{Hongjiang Chen}$^{2}$,
\name{Heyuan Huang}$^{2}$,
\name{Jihong Wang}$^{2}$,
\name{Jiaxin Yin}$^{2}$, 
\name{Jingwei Yu}$^{1}$\footnotemark[2]~, \\
\name{Junwei Liao}$^{1}$\footnotemark[2]~,
\name{Qiuying Peng}$^{2}$,
\name{Xingyu Lou}$^{2}$\thanks{Corresponding authors.}~,
\name{Jun Wang}$^{2}$,
\name{Weiwen Liu}$^{1}$\footnotemark[3]~, \\
\name{Zhuosheng Zhang}$^{1}$\footnotemark[3]~, 
\name{Weinan Zhang}$^{1}$
\vspace{+0.1cm}\\
$^{1}$Shanghai Jiao Tong University \quad
$^{2}$OPPO Research Institute\\
\texttt{lining01@sjtu.edu.cn}\quad\texttt{linqiqiang1@oppo.com} \\
\texttt{louxingyu@oppo.com}\quad\texttt{\{wwliu, zhangzs\}@sjtu.edu.cn}
}

\begin{document}

\maketitle

\begin{abstract}
With the advancements in hardware, software, and large language model technologies, the interaction between humans and operating systems has evolved from the command-line interface to the rapidly emerging AI agent interactions. 
Building an operating system (OS) agent capable of executing user instructions and faithfully following user desires is becoming a reality. In this technical report, we present ColorAgent, an OS agent designed to engage in long-horizon, robust interactions with the environment while also enabling personalized and proactive user interaction. To enable long-horizon interactions with the environment, we enhance the model’s capabilities through step-wise reinforcement learning and self-evolving training, while also developing a tailored multi-agent framework that ensures generality, consistency, and robustness. In terms of user interaction, we explore personalized user intent recognition and proactive engagement, positioning the OS agent not merely as an automation tool but as a warm, collaborative partner. We evaluate ColorAgent on the AndroidWorld and AndroidLab benchmarks, achieving success rates of 77.2\% and 50.7\%, respectively, establishing a new state of the art. Nonetheless, we note that current benchmarks are insufficient for a comprehensive evaluation of OS agents and propose further exploring directions in future work, particularly in the areas of evaluation paradigms, agent collaboration, and security.
\end{abstract}

\section{Introduction}

Over the past decades, the way humans interact with operating systems (OS) has undergone continuous transformation, from command-line~\citep{stephenson1999beginning} to graphical user interface (GUI)~\citep{toby2001expgui}, and more recently to voice-assisted~\citep{hoy2018alexa} and AI-assisted~\citep{mei2024aios} paradigms. These shifts point toward a future where the operating system itself becomes an intelligent mediator of user intent, embodied in what we call an OS Agent. An OS Agent is envisioned as a persistent, context-aware system that not only understands user instructions but can also autonomously orchestrate device functionalities to accomplish complex goals, towards a super-intelligent AI assistant bridging human and digital devices~\citep{hu2025agents}.

Recent advancements in (multimodal) large language models ((M)LLMs) and agentic frameworks have accelerated this vision, with numerous works building agents to complete user-specified tasks by operating devices autonomously. Various studies~\citep{hong2024cogagent, gou2025uground, wang2025ui, wu2025osatlas, wang2025opencua, gu2025ui} focus on building end-to-end models with enhanced reasoning, perception, and grounding capabilities to support precise operations when completing user tasks. Meanwhile, popular agent frameworks have been proposed~\citep{gur2024a, li2024appagentv2advancedagent, agashe2025agents2compositionalgeneralistspecialist, ye2025mobile}, leveraging the cutting-edge (M)LLMs to decompose user tasks into structured decision-making pipelines.

Despite significant progress in automating task execution, existing agents are largely positioned as passive ``task executors'' rather than ``interactive partners''. In contrast, an OS Agent focuses more on real-world, context-aware interactions, adapting not only to the environment but also to dynamic user needs.
When interacting with the environment, the OS agent also acts as a task executor, but can stably interact with the environment in long-horizon and complex scenarios, which is not fully explored in previous works~\citep{liu2025verigui,song2025colorbenchbenchmarkingmobileagents}.
On the other hand, the OS Agent should be able to align with human intentions, including navigating ambiguous intentions and proactively interacting with the users. For instance, clarify a vague ``\textit{Order a hamburger}'' request with a specific flavor by asking the user or reasoning according to the user's historical interests instead of acting randomly. Unlike traditional agents that execute tasks in isolation, an OS Agent must continuously adjust its behavior in response to evolving user input and contextual shifts.

\begin{figure}[t]
    \centering
    \vspace{-15pt}
    \includegraphics[width=1\linewidth]{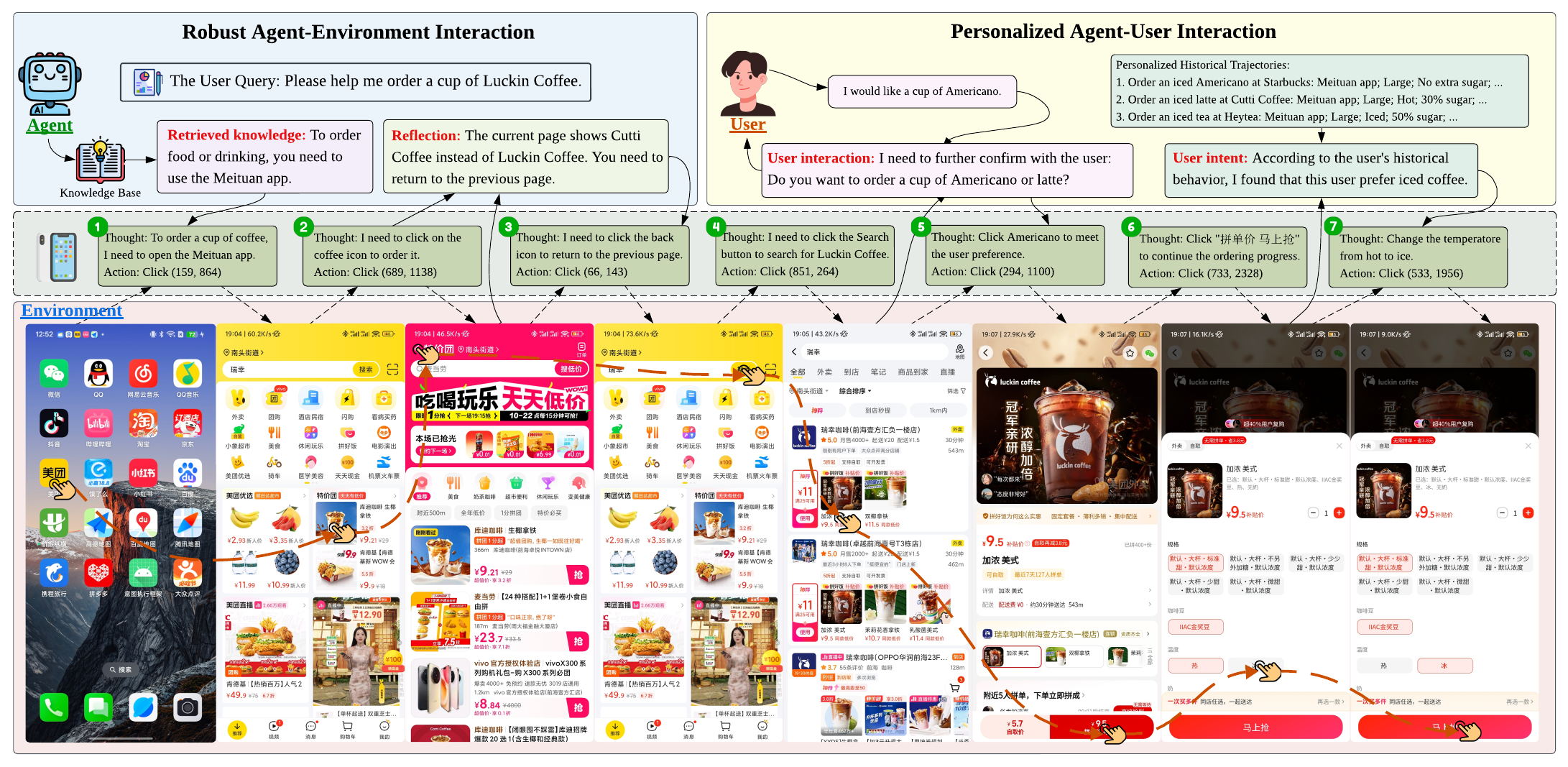}
    \vspace{-15pt}
    \caption{An overview of how the OS agent interacts with both the environment and the user.}
    \vspace{-15pt}
    \label{fig:intro}
\end{figure}

In this technical report, we propose ColorAgent, an OS agent designed to engage in long-horizon, robust interactions with the environment while also enabling personalized and proactive user interaction, as exemplified in Figure \ref{fig:intro}. ColorAgent is built as a GUI agent on mobile operating systems, which receives user instructions and operates mobile devices. To achieve robust environmental interaction, we present a tailored training paradigm and a sophisticated multi-agent framework. Furthermore, to foster a warm, collaborative partnership with the user, we explore novel mechanisms for personalized and proactive user interaction. The primary contributions of our work are as follows:
\begin{itemize}[leftmargin=15pt]
    \item \textbf{Tailored Training Paradigm and Agent Framework.} To support long-horizon and robust interactions in dynamic mobile environments, we introduce a twofold enhancement. At the model level, we adopt step-wise reinforcement learning and self-evolving training to refine the agent's grounding, perception, and reasoning abilities, enabling it to adapt to complex and dynamic GUI environments. At the framework level, we construct a multi-agent architecture that decouples task management from execution, integrates retrieval-augmented knowledge for better generalization, and employs hierarchical reflection to detect and recover from errors. 
    Our twofold enhancement enables ColorAgent to accurately perceive and manipulate the environment,  while also maintaining generalization, consistency, and stability across complex tasks.
    \item \textbf{Personalized and Proactive User Interaction.}
    Beyond task execution, we emphasize the role of ColorAgent as a collaborative partner that aligns closely with human intent. To achieve this, we explore two complementary approaches. First, when additional user memory (e.g., histories, profiles, or preferences) is available, the agent can leverage explicit signals from past trajectories and implicit cues from user habits to personalize its behavior. Second, when no additional user memory is available, the agent proactively engages the user to clarify ambiguous intentions or incomplete instructions. It learns when to trust its environment and when to query the user, ensuring alignment through active dialogue. These mechanisms allow ColorAgent to move beyond a cold, utilitarian tool and evolve toward a warm, interactive partner.
\end{itemize}

We evaluate the autonomous task execution capability of ColorAgent on two widely adopted dynamic Android benchmarks, AndroidWorld~\citep{rawles2025androidworld} and AndroidLab~\citep{xu2024androidlabtrainingsystematicbenchmarking}. Experimental results demonstrate that ColorAgent achieves state-of-the-art (SOTA) performance, attaining success rates of 77.2\% and 50.7\%, respectively. In addition to assessing the agent's interaction with the environment, we also evaluate its human-agent interaction capabilities in specific scenarios. Results on MobileIAR~\citep{wu2025quick} and VeriOS-Bench~\citep{wu2025verios} indicate that our approach outperforms all baseline models, achieving performance scores of 58.66\% and 68.98\%, respectively. 
Despite these promising results, a comprehensive benchmark that captures the full spectrum of OS agent capabilities remains absent. 
To build a truly practical OS Agent, we advocate for the research community to move beyond isolated success-rate metrics and adopt more holistic evaluation paradigms that better reflect real-world complexity. 
Toward this goal, we outline several key research directions for advancing the field, including: (i) more comprehensive and nuanced evaluation protocols, (ii) better multi-agent collaboration frameworks, and (iii) robust mechanisms for safe and reliable execution.


\section{Model Training}

\begin{figure}[t]
    \centering
    \vspace{-15pt}
    \includegraphics[width=1\linewidth]{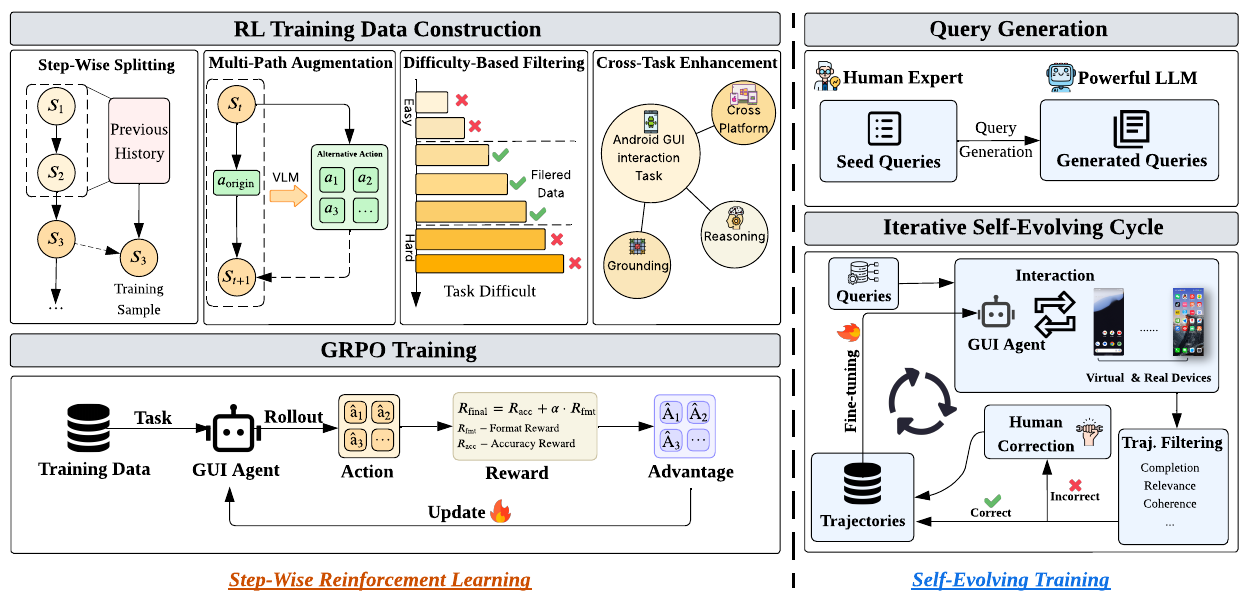}
    \caption{Our two-stage training framework for developing a powerful GUI model. The process begins with \textbf{Step-wise Reinforcement Learning (left)} to optimize the agent's core decision-making abilities on constructed training data. It then progresses to \textbf{Self-evolving Training (right)}, a iterative loop that generates high-quality trajectory data to enable continuous improvement.}
    \label{fig:training_framework}
\end{figure}

To develop a robust GUI model with precise interaction and deep reasoning capabilities as the backbone of ColorAgent, we utilize a two-stage progressive training paradigm. It addresses key challenges like perception ambiguity and action grounding by sequentially enhancing the model, moving from single-step optimization to iterative, data-driven improvement. 
The overall paradigm is illustrated in Figure \ref{fig:training_framework} and consists of two sequential stages:

\begin{itemize}[leftmargin=15pt]
\item \textbf{Stage I: Step-wise reinforcement learning (§\ref{sec:step-wise-rl})} focuses on optimizing single-step decision-making. It employs reinforcement learning with tailored rewards to significantly improve the model's reasoning and action accuracy within complex GUI contexts.
\item \textbf{Stage II: Self-evolving training (§\ref{sec:self_evolving})} tackles the data bottleneck by creating a self-sustaining data generation loop. 
This stage enables the model to automatically produce and refine high-quality interaction data, forming a reinforcing cycle that reduces the reliance on heavy manual annotation.
\end{itemize}

\subsection{Stage I: Step-Wise Reinforcement Learning} \label{sec:step-wise-rl}
Stage I employs step-wise sample-based reinforcement learning (RL) to enable the model to explore optimal actions given observations—specifically, historical interaction records and current GUI screenshots. This stage prioritizes equipping the model with effective decision-making patterns through exploration and feedback, instead of adhering to pre-defined "correct" reasoning patterns.

We define two core objectives for this stage: first, exploring robust single-step reasoning processes to improve decision accuracy across diverse GUI contexts; second, ensuring the model’s outputs are both interpretable and executable, which lays a foundation for subsequent trajectory-level optimization.
To accomplish these objectives, we designed a customized RL framework integrating two key components: (1) adaptive data construction, which simulates real-world GUI complexities; (2)  rule-based rewards, which provide clear optimization signals.

\subsubsection{Data Construction}
\label{stepwise-rl-data-construction}

To ensure the diversity and representativeness of training data—critical for adapting to complex GUI environments—we leverage seven public GUI interaction datasets: \textit{Aguvis}~\citep{Aguvis}, \textit{AITW}~\citep{AITW}, \textit{AITZ}~\citep{AITZ}, \textit{AMEX}~\citep{AMEX}, \textit{AndroidControl-High}~\citep{AndroidControl}, \textit{GUIAct}~\citep{GUIAct}, and \textit{GUI-Odyssey}~\citep{GUI-Odyssey}. These datasets collectively provide large-scale offline GUI interaction trajectories. Each trajectory can be formally defined as $\{I,s_0,a_0,\cdots,s_T,a_T\}$, where $I$ denotes the user task instruction (e.g., "adjust screen brightness"), $s_t$ represents the GUI screenshot at step $t$, and $a_t$ is the ground-truth action executed at step $t$ to advance the task.

For step-wise reinforcement learning, data quality directly influences the model’s ability to learn generalizable GUI interaction patterns. To guarantee data quality and align with RL training requirements, our data construction pipeline focuses on four essential aspects: step-wise splitting, multi-path augmentation, difficulty-based filtering, and cross-task enhancement.

\textbf{Step-Wise Splitting.} We decompose the aforementioned offline trajectories $\{I, s_0, a_0, \dots, s_T, a_T\}$ into $T$ step-wise training samples $\{I, h_{t-1}, s_t, a_t\}$, where $h_{t-1}$ is the operational history in the previous $t-1$ steps. It is generated by Qwen2.5-VL-72B~\citep{bai2025qwen2} that summarizes action descriptions from each step in the historical operations. This decomposition preserves contextual dependencies between consecutive interaction steps—ensuring the model retains memory of prior actions, while enabling RL to optimize intermediate decision-making (rather than only final trajectory outcomes), which is critical for learning fine-grained GUI operation logic.

\textbf{Multi-Path Augmentation.} Traditional GUI interaction frameworks~\citep{luo2025guir1generalistr1style, lu2025uir1enhancingefficientaction} adopt a rigid single-action annotation paradigm: they treat the pre-annotated action as the sole correct target, implicitly penalizing other valid alternatives. This design fails to align with real-world GUI interaction characteristics—where multiple distinct action paths often achieve the same task goal, driven by factors such as interface redundancy (e.g., dual-function control elements) and operation habits (e.g., gesture vs. command input), etc.
Typical examples of such multi-path validity are widespread in practical use:
\begin{itemize}[leftmargin=15pt]
\item Launching a target app: either by scrolling through the app drawer to click the app icon, or by directly invoking the predefined \texttt{open\_app} function.
\item Navigating back: either by tapping the on-screen \texttt{"Back"} arrow in the app interface, or by triggering the \texttt{system\_button("Back")} command.
\end{itemize}
To address this mismatch, we use Qwen2.5-VL-72B to identify steps with multiple valid actions, expand annotations to include all verified alternatives, and assign them equal reward weight in training. This directly mitigates the bias of over-penalizing valid variants.
While multi-path labeling may slightly compromise performance on static benchmarks (where only one ``canonica'' action is accepted), it substantially boosts generalization in real-world scenarios.

\textbf{Difficulty-Based Filtering.}
To ensure that the training process focuses on informative and valuable samples, we introduce a difficulty-based filtering strategy. Specifically, we employ Qwen2.5-VL-72B to perform inference on each training sample, conducting eight independent runs per instance. The inference performance is quantified by the number of correct predictions \( c \in [0, 8] \). The samples with either perfect predictions (\( c = 8 \)) or complete failures (\( c = 0 \)) contribute little to effective model learning. To address this imbalance, we discard all samples with \( c = 8 \) and partially remove those with \( c = 0 \). This filtering process yields a more balanced and informative dataset, thereby enhancing the overall learning efficacy and generalization performance of the model.

\textbf{Cross-Task Enhancement.}
To mitigate the risk of overfitting and to bolster reasoning capabilities, the seven datasets we selected contain not only GUI interaction trajectories on mobile devices, but also data from computer and web platforms. Additionally, we incorporate two types of auxiliary tasks: (1) spatial localization-focused grounding tasks from AndroidControl-Low~\citep{AndroidControl}, which help the model accurately map the action intent to GUI element positions (e.g., locating a ``back button'' in crowded interfaces), and (2) geometric math tasks from Geometry3K~\citep{Geometry3K} to stengthen the GUI model’s general reasoning capabilities. 

\subsubsection{Group Relative Policy Optimization}

\label{sec:grpo}

We conduct step-wise reinforcement learning with the Group Relative Policy Optimization (GRPO)~\citep{shao2024deepseekmath} algorithm. GRPO estimates the relative advantage values by sampling groups of responses, eliminating the need for a separate critic model for value estimation, providing a lightweight and stable alternative to Proximal Policy Optimization (PPO)~\citep{schulman2017proximal}.

For each input sample by combining the user instruction $I$, current screenshot $s_t$, and history interactions $h_{t-1}$, the model generates $N$ candidate responses \( O = \{o_1, o_2, \ldots, o_N\} \). 
Each response is evaluated by a rule-based reward function, yielding rewards $\{r_1, r_2, \ldots, r_N\}$.
The estimated relative advantage  $\hat{A}_i$ of the $i$-th response is computed as:
\begin{equation}
    \hat{A}_i = \frac{r_i - \text{mean}(\{r_1, \ldots, r_N\})}{\text{std}(\{r_1, \ldots, r_N\}) }.
\end{equation}
 This normalization enables comparison of responses within their group, capturing nuanced quality differences without an absolute reward scale.

After estimating the relative advantages, the policy is updated by maximizing a clipped surrogate objective, regularized by Kullback-Leibler (KL) divergence to prevent abrupt policy shifts:
\begin{equation}
    \mathcal{J}_{\text{GRPO}}(\theta) = \mathbb{E} \left[ \frac{1}{N} \sum_{i=1}^N \min \left( \frac{\pi_\theta(o_i)}{\pi_{\theta_{\text{old}}}(o_i)} \hat{A}_i, \text{clip}\left( \frac{\pi_\theta(o_i)}{\pi_{\theta_{\text{old}}}(o_i)}, 1\pm\epsilon\right) \hat{A}_i \right) - \beta D_{\text{KL}}(\pi_\theta || \pi_{\text{ref}}) \right],
\end{equation}
where \( \pi_\theta \) and \( \pi_{\theta_{\text{old}}} \) denote the current and previous policies, \( \pi_{ref} \) is the reference model, \( \epsilon \) (clipping threshold) and \( \beta \) (KL coefficient) stabilize training, ensuring smooth policy evolution across diverse GUI environments.

\subsubsection{Rule-Based Rewards}
For each response $o_i \in O$ generated by the model, it includes three parts: a thought containing the reasoning process (e.g., ``To find the target item, I need ...''), an action summary to describe the action in natural language (e.g., ``Click the search button ...''), and an executable action in JSON format. The action is composed of an action type and its possible parameters. Detailed action space is listed in Appendix \ref{app:action-space}.

To evaluate the responses, we apply a rule-based reward consisting of two aspects: the \textbf{format reward} and the \textbf{accuracy reward}. The format reward $R_{fmt}$ assesses whether the response includes all three parts with a pre-defined format, and whether the action in JSON format can be correctly parsed. It is defined as: 
\begin{equation}
    R_{\text{fmt}} =
    \begin{cases}
    1 & \text{if all parts are correctly formatted,} \\
    0 & \text{otherwise.}
    \end{cases}
\end{equation}

The accuracy reward \(R_{\text{acc}}\) measures whether both the predicted action type and its parameters are correct, defined as:
\begin{equation}
R_{\text{acc}} = 
\begin{cases} 
1 & \text{if } R_{\text{type}} = 1 \land R_{\text{params}} = 1, \\
0 & \text{otherwise.}
\end{cases}
\end{equation}

Here, \( R_{\text{type}} = 1 \) if the action type matches the ground truth, and \( R_{\text{params}} = 1 \) if the action's parameters are correct. Specifically, coordinate-based actions are correct when the predicted coordinates fall within the ground-truth bounding box; text-related actions are validated by semantic similarity between the predicted and reference texts (e.g., F1 score); and swipe actions are correct when the predicted swipe direction matches the ground truth.

The final reward combines both the accuracy and format adherence:
\begin{equation}
    R_{\text{final}} = R_{\text{acc}} + \alpha \cdot R_{\text{fmt}},
\end{equation}
where $\alpha=0.2$ is used to balance the two terms while maintaining the emphasis on action accuracy.

\subsection{Stage II: Self-Evolving Training}
\label{sec:self_evolving}

To scale up trajectory data and reduce the heavy reliance on manual annotation, we propose a self-evolving training pipeline that establishes a robust reinforcing cycle of ``data generation → model optimization → higher-quality data generation'' based on the model refined in Stage I. As illustrated in the right panel of Figure~\ref{fig:training_framework}, this iterative process begins with the \textbf{query generation} phase, where a dual-source strategy integrates human expertise and language model expansion. The generated queries are then processed through an \textbf{iterative self-evolving cycle}, which systematically generates, evaluates, and refines interaction data to continuously enhance model performance. Through this continual loop, the model progressively enhances its capabilities while breaking free from the limitations of static, manually annotated datasets.

\subsubsection{High-Quality Query Generation}

To produce meaningful interaction trajectories, generating realistic, diverse, and task-relevant queries is essential. Our approach achieves this through a dual-source method that synergizes human expertise with a powerful language model. First, we utilize domain experts to craft a set of high-quality seed queries. This manual step ensures the queries are practical, executable, and reflect real-world user interactions, establishing a strong foundation of ecological validity. Subsequently, we employ a powerful large language model, DeepSeek-R1~\citep{deepseekai2025deepseekr1incentivizingreasoningcapability}, to expand upon these seeds, generating a wide array of syntactically and semantically varied queries. This model-driven expansion is crucial for ensuring comprehensive coverage of possible user interactions, including challenging corner cases that human experts may not intuitively consider.

\subsubsection{Iterative Self-Evolving Cycle}
After generating high-quality queries, the training process enters an iterative self-evolving cycle designed to continuously refine both the data and the model itself. Each iteration of the self-evolving cycle consists of three sequential stages: trajectory rollout, trajectory filtering, and fine-tuning.

\textbf{Trajectory Rollout.} 
Using the pre-generated query pool, the current-stage model interacts with a dual-environment setup, combining Android virtual environments and ColorOS physical devices, to produce step-by-step interaction trajectories. 
For each query, the rollout is repeated multiple times to capture diverse interaction paths by varying initial conditions (starting pages) and decision-making strategies (a high temperature), which reflect the fluctuations of real-world user behavior and enrich the data diversity.

\textbf{Trajectory Filtering.} 
To ensure that only high-quality data enters training, we design a stringent filtering module composed of multiple specialized discriminators. Each discriminator evaluates a specific dimension of trajectory quality, such as task completion, action validity, path relevance, reasoning coherence, and other relevant criteria, which is detailed in Appendix \ref{app:discriminator}. The discriminators collectively cover a broad spectrum of critical evaluation aspects for trajectory validity. 

Trajectories that pass all discriminators are retained for further fine-tuning. Those identified as incorrect trajectories (i.e., failing any of the discriminators) will be subjected to manual evaluation to locate and correct the flawed steps, and then be integrated into the fine-tuning dataset. These corrected trajectories are particularly valuable, as they address specific gaps in the model’s decision-making and provide targeted training signals.

\textbf{Fine-Tuning.} 
The filtered and corrected trajectories are utilized for supervised fine-tuning on the current model. This fine-tuning step enhances the model’s ability to generate accurate and coherent interaction trajectories, effectively preparing it for the next iteration of the self-evolving cycle. This iterative loop ensures that both the model and the quality of the data are mutually enhanced over time.

\section{Agent Framework}

\subsection{Why the Single Agent Failed?}
\label{limitation of single agent}

\begin{wrapfigure}{r}{0.3\textwidth}  
    \centering
    \includegraphics[width=0.25\textwidth]{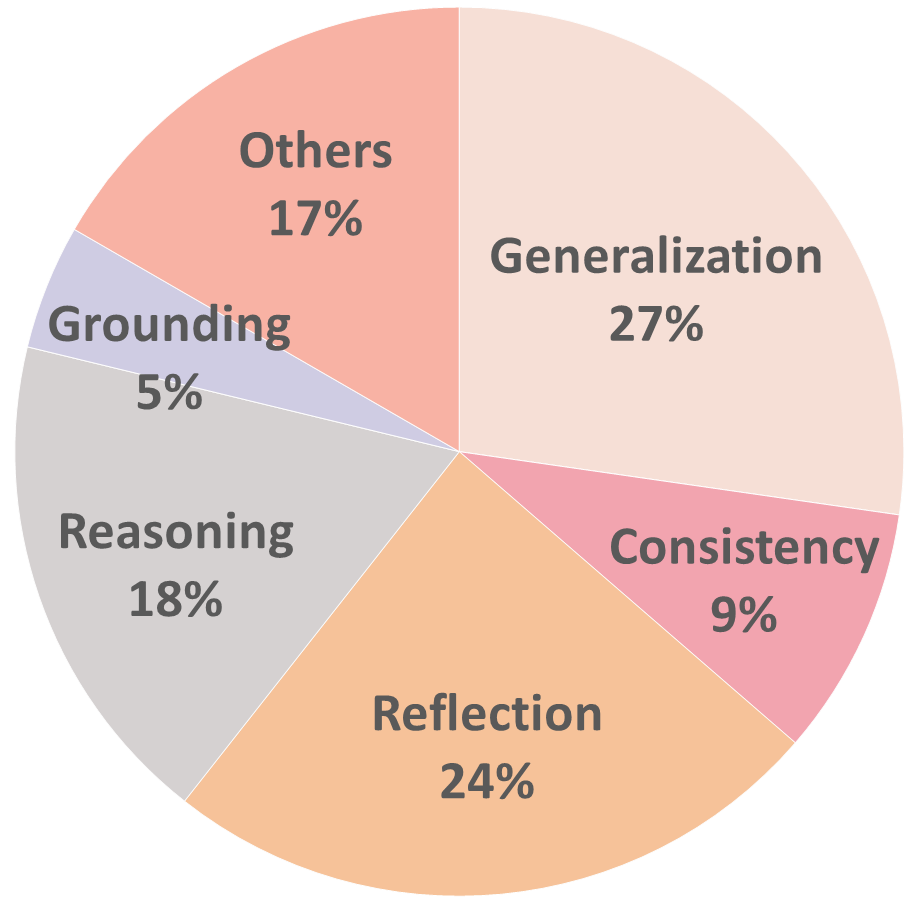}
    \caption{ Error type distribution of the single agent.}
    \vspace{-20pt}
    \label{fig:error_pie}
\end{wrapfigure}

Although the single GUI agent (directly built upon the GUI model) has made significant advancements in capabilities such as grounding and reasoning through extensive post-training on large-scale datasets, a substantial gap remains before they can be robustly deployed in the high variability and uncertainty of real-world mobile environments. To diagnose the root causes of these failures, we analyzed the error distribution of a single agent on the AndroidWorld~\citep{rawles2025androidworld} benchmark. Our analysis, summarized in Figure \ref{fig:error_pie}, reveals a clear pattern: more than half of the failures stem from a lack of three core capabilities: generalization, reflection, and consistency. This data-driven insight underscores the fundamental limitations of a single-agent paradigm and motivates our adoption of a multi-agent framework.

\subsubsection{Limited Generalization}\label{lim:gemeralization}

A single GUI agent, optimized on large-scale training datasets, can achieve strong in-domain performance but easily fails in out-of-domain environments or tasks~\citep{wu2025gem}. For example, when performing a search operation, an agent knows to first click the search box and then enter the target content. However, a minor UI variation, such as a search box requiring two consecutive clicks to activate, can trap the agent in a repetitive and unproductive loop. This issue is compounded by the agent's lack of self-evolution: It is unable to learn from past trajectories to reuse prior experiences or avoid past mistakes. This inability to adapt to minor UI variations or learn from new experiences highlights a fundamental flaw: A single agent lacks a mechanism to dynamically incorporate external knowledge or evolve its strategy.

\subsubsection{Inconsistency and Lack of Memory}\label{lim:inconsistency}

In real-world scenarios, there are often complex, long-horizon tasks that involve compositional goals and intricate dependencies~\citep{guo2025atomic}. It's challenging for a single agent to maintain consistency and transfer information across long-term task execution. Take the task ``Find the price of OPPO Find X9 in shopping apps A, B, and C, and add the one with the lowest price to the cart'' as an example. The agent may exhibit flawed behavior, such as ignoring one of the apps, or fail entirely due to its inability to retain key information—the phone price in each app—throughout its execution. This example illustrates that without dedicated mechanisms for task decomposition, progress tracking, and cross-step memory management, a single agent is inherently incapable of handling the demands of compositional, long-horizon tasks.

\subsubsection{Difficulty in Error Recovery}\label{lim:reflection}

When operating a mobile device, the agent may make various mistakes, such as clicking the wrong icon and entering an unintended page, or forgetting to activate the input field before entering text. If these initial errors are not corrected promptly, the agent may engage in meaningless exploration on the wrong page or fall into repetitive actions. Ultimately, the agent may mistakenly terminate the task, overlooking several critical steps. 

Making mistakes is inevitable, even for the most powerful GUI agents. The key lies in how to detect errors and recover from them. While some work has attempted to enable the single agent with self-correction capabilities~\citep{wu2025guireflectionempoweringmultimodalgui, wanyan2025lookleapguicriticr1model}, this is typically limited to individual steps, neglecting longer-term error recovery. Instead, enabling reflection at different scales of task execution can effectively enhance the robustness, thereby improving the applicability of GUI agents in real-world scenarios. After all, the cost of recovering from errors is far lower than the significant loss caused by task failure.

\begin{figure}[t]
    \centering
    \includegraphics[width=1\linewidth]{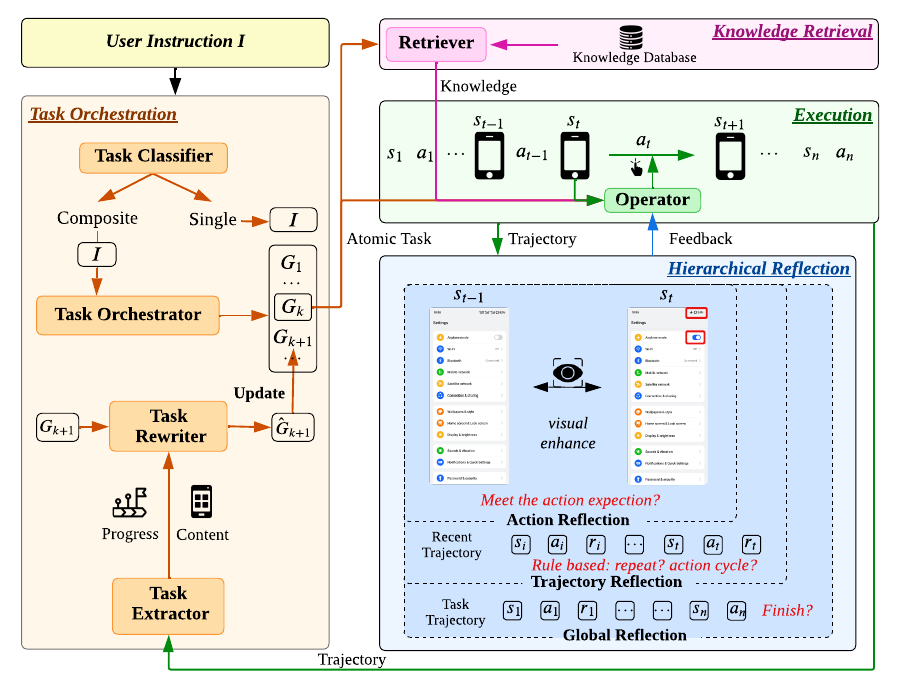}
    \caption{ The architecture of our multi-agent framework. It consists of a central Execution module supported by three key components: the Task Orchestration module decomposes complex user instructions into manageable atomic tasks and handles memory transfer; the Knowledge Retrieval module provides relevant external information; and the Hierarchical Reflection module delivers multi-level feedback for error correction.}
    \label{fig:agent_framework}
\end{figure}

\subsection{Multi-agent Framework}

To address the fundamental limitations of a single agent detailed above, we construct a multi-agent framework\footnote{The implementation of our agent framework is available at \url{https://github.com/MadeAgents/mobile-use}.} designed to significantly enhance its capabilities. As illustrated in Figure \ref{fig:agent_framework}, our framework is built around a central execution module, which is augmented by three specialized, complementary modules, each directly targeting a core deficiency identified in Section~\ref{limitation of single agent}:
\begin{itemize}[leftmargin=15pt]
\item To combat limited generalization (§~\ref{lim:gemeralization}), a \textbf{Knowledge Retrieval} module provides dynamic access to an external knowledge base.
\item To overcome inconsistency and lack of memory (§~\ref{lim:inconsistency}), a \textbf{Task Orchestration} module decomposes complex goals and manages information flow across steps.
\item To mitigate difficulty in error recovery (§~\ref{lim:reflection}), a \textbf{Hierarchical Reflection} module enables multi-level error detection and correction.
\end{itemize}

Together, these components form a synergistic system where the execution module's core abilities are amplified, leading to more robust, consistent, and intelligent task completion. The following subsections will detail each of these components.

\subsubsection{Knowledge Augmentation with Retrieval}

To enhance the agent's adaptability to a wide range of tasks, we introduce a knowledge retrieval module that provides task-specific knowledge. Given a user instruction $I$ or a decomposed atomic task $G_k$ as a query $q$, a retriever $R$ searches a diverse knowledge database to find the most pertinent information. This knowledge base can contain manually constructed experiences, web-scraped content, or insights derived from historical trajectories. This entire process can be formally represented as:
\begin{equation}
    K_q =R(q, D),
\end{equation}
where $K_q$ is the retrieved knowledge with query $q$, and $D$ is the knowledge database.

During task execution, the retrieved knowledge is provided to the execution module, equipping it with prior knowledge about the current environment and task, improving its reasoning ability and reducing errors. For example, when executing the task ``Find my high-priority tasks in the Tasks app'', the knowledge retrieved from the knowledge base, ``In the Task app, red represents high priority'', will play a critical role in completing the task.

\subsubsection{Task Management and Memory Transfer with Task Orchestration}

The task orchestration module serves as the central hub for managing the entire task. Given a user instruction $I$, a task classifier $TC$ first determines whether the current task is complex and decomposable (e.g., tasks that involve multiple sub-goals, cross-app interactions, or information transfer). If so, a task orchestrator $TO$ will decompose it into a sequence of manageable atomic tasks:
\begin{equation}
    \{G_1, G_2, \dots\} = TC(I), \quad \text{if}\,\, TO(I) = \text{Composite},
\end{equation}
where $\{G_1, G_2, \dots\}$ are a series of atomic tasks. Unlike traditional planner~\cite{NEURIPS2024_0520537b}, which require task-specific understanding (such as decomposing the task ``open wifi'' into ``open settings, find the wifi switch, turn on wifi''), our task orchestration operates at a higher level (for example, decomposing ``turn on wifi, increase phone brightness'' into two atomic tasks: ``turn on wifi'' and ``increase phone brightness''), preserving the natural structure of the user's intent. 

The primary challenge in executing this sequence of atomic tasks is maintaining information flow and context. To address this, we introduce a memory transfer mechanism. After the execution module completes an atomic task $G_k$, a task extractor $TE$ distills the critical information from the trajectory for completing $G_k$. A task rewriter $TR$ then integrates this content into the next atomic task $G_{k+1}$ to form an updated, context-aware instruction $\hat{G}_{k+1}$:
\begin{equation}
    \hat{G}_{k+1} = TR(G_{k+1}, TE(\{s_1, a_1, \dots, s_n, a_n\})),
\end{equation}
where $\{s_1, a_1, \dots, s_n, a_n\}$ are the trajectory (screenshots and actions) when completing $G_k$.
For example, after completing the first atomic task ``View the expenses from expenses.jpg in Simple Gallery Pro.'' of the user instruction ``Add the expenses from expenses.jpg in Simple Gallery Pro to pro expense'', the task extractor captures the contents of ``expenses.jpg''.  After that, the task rewriter will reformulate the second atomic task into ``Add the following expenses to pro expense: xxx''. This process allows the agent to handle complex dependencies and robustly execute long-horizon tasks.

\subsubsection{Error Detection and Recovery with Hierarchical Reflection}

Inherited from MobileUse \citep{li2025mobileuse}, we incorporate a hierarchical reflection module to improve the resilience of autonomous mobile systems by supporting error identification and correction across various stages of task execution. It consists of three major components:
\begin{itemize}[leftmargin=15pt]
    \item \textbf{Action Reflector} is responsible for real-time monitoring of individual actions. For every step, it inspects the screenshots captured before and after the action execution to determine whether the intended outcome is achieved. When issues such as grounding errors, visual misperceptions, or interface misinterpretations arise, the Action Reflector produces diagnostic feedback that highlights the problem and its potential cause. The execution module then leverages this feedback to adjust its subsequent behavior.
    \item \textbf{Trajectory Reflector} oversees short sequences of actions to track ongoing progress.  It analyzes recent steps (typically the last 3–5 actions) together with the corresponding action-level reflections to verify whether the execution remains coherent and aligned with the task objective. Once inconsistencies are found, it delivers corrective feedback to help the execution module refine its trajectory and continue moving effectively toward the final goal.
    \item \textbf{Global Reflector} provides an overall assessment at the task level. Triggered only once a task has reached a tentative endpoint, it reviews the full sequence of actions along with the latest screenshots to decide whether the original instruction has been satisfactorily completed. If the task is judged incomplete, the Global Reflector supplies feedback that prompts the execution module to resume and complete the remaining steps in subsequent iterations.
\end{itemize}

By integrating these three levels of reflection, the hierarchical module equips the agent with the ability to capture and correct errors at different granularities. Each reflector delivers targeted feedback that guides the execution module in revising strategies, reducing unnecessary repetition, and ensuring continuous alignment with user instructions throughout the execution process.

\section{Experiments}

\subsection{Experimental Settings}
\textbf{Benchmarks.} We evaluate the proposed training strategies and agent framework on two widely used mobile benchmarks, AndroidWorld~\citep{rawles2025androidworld} and AndroidLab~\citep{xu2024androidlabtrainingsystematicbenchmarking}, to assess their effectiveness in enhancing environment interaction. AndroidWorld comprises 116 tasks drawn from 20 mobile applications, while AndroidLab includes 138 tasks across 9 applications. In both benchmarks, the agent is provided with natural language instructions and required to dynamically interact with a pre-initialized Android environment, with task success determined according to benchmark-defined rules. In addition, several issues identified in the original benchmarks were fixed, with details provided in Appendix \ref{app:fix-benchmark}.

\textbf{Implementation Details.} For step-wise reinforcement learning, we employ Qwen2.5-VL-72B~\citep{bai2025qwen2} and GUI-Owl-32B~\citep{ye2025mobile} as our base models, utilizing an implementation of GRPO adapted from the Verl framework~\citep{sheng2024hybridflow}. We utilize the full-parameter configuration \(8 \times 8\) A800 GPUs across 2 epochs with a learning rate of \(1 \times 10^{-6}\). The rollout number is set to 5.

\subsection{Main Results}

\begin{table}[htbp]
\centering
\resizebox{\linewidth}{!}{
\begin{tabular}{lcc}
\toprule
\textbf{Methods} & \textbf{AndroidWorld (SR)} & \textbf{AndroidLab (SR)} \\
\midrule
\textit{Proprietary Models} & & \\
GPT-4o-2024-11-20~\citep{achiam2023gpt} & 34.5 & 31.2 \\
Claude-Sonnet-4-20250514-thinking~\citep{claude4} & 41.0 & 40.6 \\
UI-TARS-1.5~\citep{qin2025ui} & 64.2 & 38.3 \\
MobileRL~\citep{xu2025mobilerl} & 75.8 & 46.8 \\
\midrule
\textit{Open Models} & & \\
Qwen2.5-VL-7B-Instruct~\citep{bai2025qwen2} & 27.6 & 10.1 \\
GLM-4.1V-9B-Thinking~\citep{hong2025glm} & 41.7 & 24.6 \\
UI-TARS-7B~\citep{qin2025ui} & 33.0 & 32.6 \\
V-Droid~\citep{dai2025advancingmobileguiagents} & 59.5 & 38.3 \\
UI-Venus~\citep{gu2025ui} & 65.9 & - \\
GUI-Owl-7B~\citep{ye2025mobile} & 66.4 & 42.8 \\
\midrule
\textit{Frameworks} & & \\
MobileUse~\citep{li2025mobileuse} & 62.9 & 44.2 \\
Mobile-Agent-v3~\citep{ye2025mobile} & 73.3 & - \\
\midrule
Qwen2.5-VL-72B-Instruct~\citep{bai2025qwen2} & 35.0 &  31.9 \\
\quad + Model Training & 64.7 & 46.4 \\
\hdashline
GUI-Owl-32B~\citep{ye2025mobile} & 54.3 & 38.4 \\
\quad + Model Training & 65.1 & 48.6 \\
\quad + Agent Feamework & \textbf{77.2} & \textbf{50.7} \\
\bottomrule
\end{tabular}
}
\caption{Success Rate (\%) on the AndroidWorld and AndroidLab benchmark.}
\label{tab:benchmark_results}
\end{table}

As shown in Table~\ref{tab:benchmark_results},  ColorAgent achieves highly competitive performance among proprietary models, open models, and frameworks on both the AndroidWorld and AndroidLab benchmarks. We attribute this success to two key, complementary factors that we analyze below: the significant performance gains from our model training paradigm and the enhanced robustness provided by our agent framework.

\textbf{Model Training Performance.} Directly applying large open-source vision-language model (VLM) yields limited success. However, with our step-wise reference learning and self-evolving training strategies, remarkable improvements are exhibited on the Qwen2.5-VL-72B-Instruct and GUI-Owl-32B models. The former achieved a huge performance improvement of 29.3\% and 14.5\%, respectively, in AndroidWorld and AndroidLab. Although GUI-Owl-32B has been trained on a large amount of GUI data, our training strategy can still bring further improvements. These results confirm that our training paradigm significantly enhances the reasoning, perception, and grounding capabilities of GUI models, achieving results that are competitive with the state-of-the-art models.

\textbf{Agent Framework Performance.} Beyond model-level improvements, incorporating the proposed Agent framework further boosts performance, particularly in complex, long-horizon tasks. Building upon GUI-Owl-32B with model training, the addition of our knowledge retrieval, task orchestration, and hierarchical reflection modules increases the success rate to 77.2\% on AndroidWorld and 50.7\% on AndroidLab. This establishes a new state of the art (SOTA) among open models and frameworks, outperforming prior systems such as MobileUse (62.9\% / 44.2\%) and Mobile-Agent-v3 (73.3\% on AndroidWorld). The results highlight the complementary nature of our contributions: while training improves the intrinsic capabilities of the base models, the agent framework ensures more consistent task execution, better error recovery, and stronger adaptability in dynamic environments.

\subsection{Further Analysis}

\begin{figure}[ht]
    \centering
    \begin{minipage}[b]{0.48\textwidth}
        \centering
        \includegraphics[width=\linewidth]{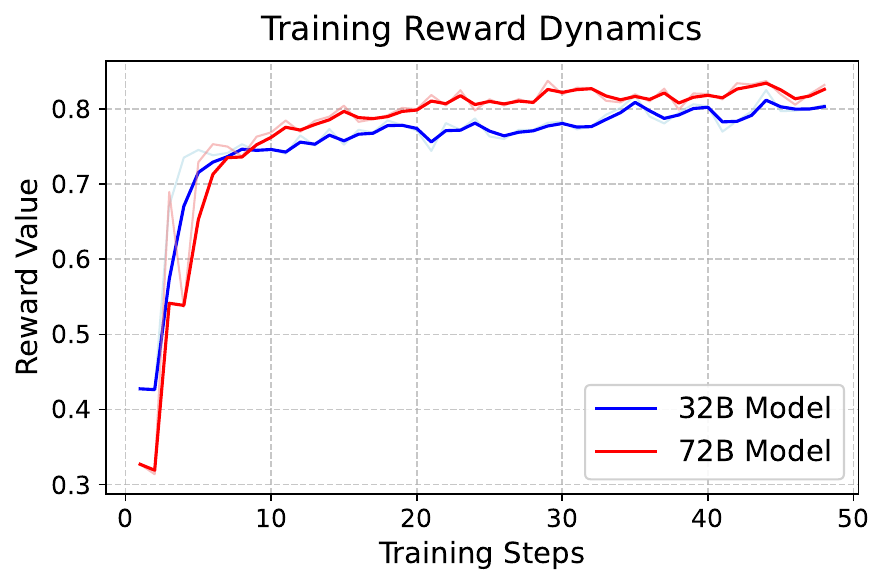}
        \vspace{-15pt}
        \caption{Training reward dynamics of different models.}
        \label{fig:training_curve}
    \end{minipage}
    \hfill
    \begin{minipage}[b]{0.48\textwidth}
        \centering
            \resizebox{\linewidth}{!}{
            \begin{tabular}{lc}
            \toprule
            Methods & AndroidWorld (SR) \\
            \midrule
            Qwen2.5-VL-72B-Instruct & 35.0 \\
            \hdashline
            \quad + Step-wise RL & 58.3 \\
            \quad + Self-evolving & 64.7 \\
            \midrule
            GUI-Owl-32B & 54.3 \\
            \hdashline
            \quad + Step-wise RL & 63.0 \\
            \quad + Self-evolving & 65.1 \\
            \hdashline
            \quad + Hierarchical Reflection & 70.3 \\
            \quad + Task Orchestration & 72.8 \\
            \quad + Knowledge Retrieval & 77.2 \\
            \bottomrule
            \end{tabular}
            }
        \vspace{5pt}
       \captionof{table}{Ablation study on the AndroidWorld benchmark.}
        \label{tab:ablation}
    \end{minipage}
\end{figure}

\textbf{Training Dynamics and the Generalization Trade-off}. The training dynamics, shown in Figure~\ref{fig:training_curve}, reveal a valuable insight into the trade-off between model capacity and generalization. While both the 32B and 72B models demonstrate successful learning with rapidly increasing rewards, the larger 72B model ultimately achieves a higher final reward, indicating a superior fit to the training data. However, this superior training-fit does not translate to better performance on the downstream benchmarks. As reported in Table~\ref{tab:benchmark_results}, the fine-tuned 32B model decisively outperforms the 72B model in the dynamic test environments. This discrepancy strongly suggests that the 72B model, despite its larger capacity, is more prone to overfitting the training data, which harms its ability to generalize to unseen scenarios. This finding underscores a critical challenge in developing GUI agents: balancing the expressive power of large models with the need for robust generalization remains an important avenue for future research.

\textbf{Ablation Study.} Table~\ref{tab:ablation} presents the ablation results on the AndroidWorld benchmark. For Qwen2.5-VL-72B-Instruct, the baseline achieves only 35.0\% success rate, while the introduction of step-wise reinforcement learning (RL) substantially improves performance to 58.3\%, and further gains are observed when applying self-evolving training, reaching 64.7\%. A similar trend is observed for GUI-Owl-32B, demonstrating the generality of our training strategies across model scales. Building upon the trained GUI-Owl-32B, we further analyze the contribution of each component in our agent framework. Hierarchical reflection enhances robustness to execution errors, improving the success rate to 70.3\%. Task orchestration, which decomposes complex instructions and manages memory, provides additional gains. Finally, the incorporation of knowledge retrieval yields the highest performance at 77.2\%, highlighting the importance of external knowledge for tackling diverse tasks. These results confirm that both the training strategies and agent-level modules contribute complementary improvements, and together they enable ColorAgent to achieve state-of-the-art performance on AndroidWorld.

\section{From Tool to Partner: Building Warm OS Agents Beyond Task Execution}

\begin{figure}[t]
    \centering
    \includegraphics[width=1\linewidth]{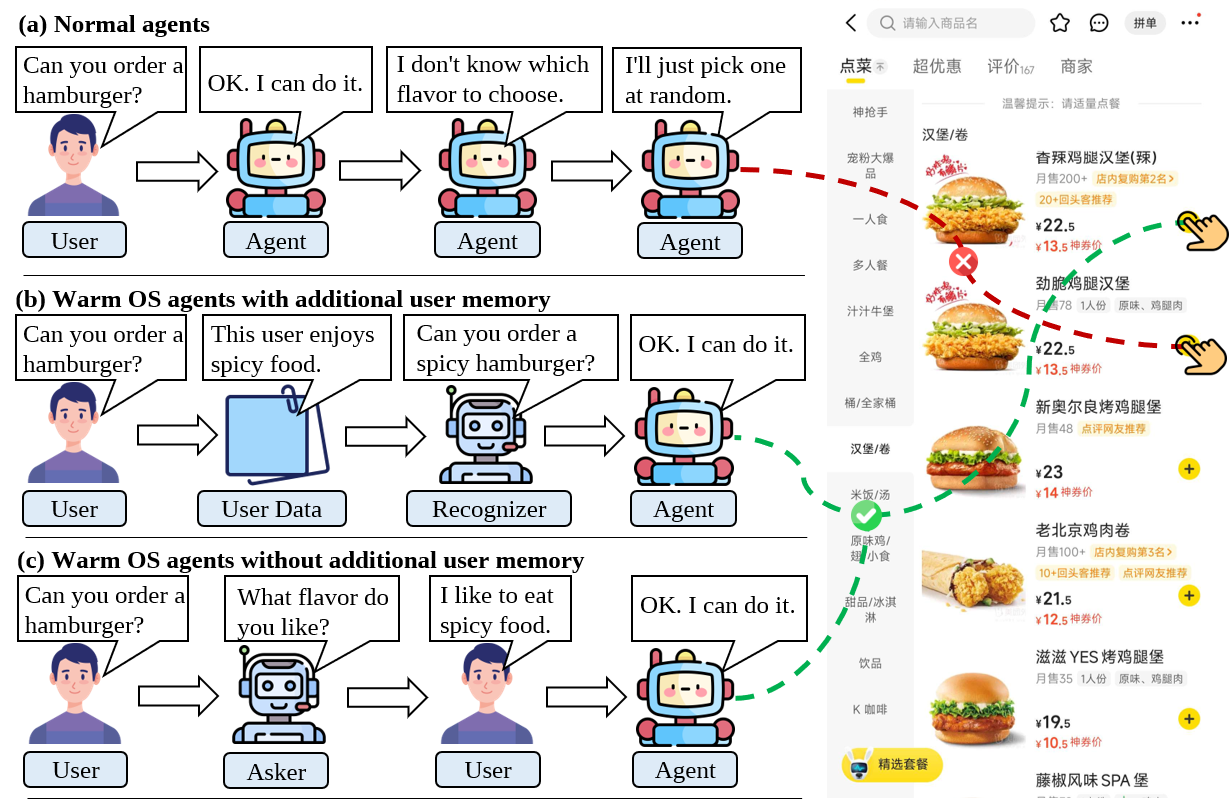}
    \caption{Comparison between a normal agent and a warm OS agent. The normal agent only follows the ordering instructions but may not select the burger type that aligns with human intent, whereas the warm OS agent accurately chooses the type of burger that the human actually wants.}
    \label{fig:warm}
\end{figure}

Although agents are becoming increasingly capable of completing tasks~\citep{ye2025mobile, wang2025ui}, focusing solely on enhancing their task-execution abilities will only result in a cold, utilitarian tool. 
To transform them into warm, collaborative partners like Jarvis in Iron Man, we must enhance their ability to understand and align with human intentions.

As shown in Figure~\ref{fig:warm}, aligning with human intentions can be approached under two task settings:
(i) with additional user memory (e.g., user interaction history, user profiles), and
(ii) without additional user memory.
In the setting with additional user memory, OS agents need to be capable of analyzing user memory to perform tasks in a personalized manner, whereas without additional user memory, OS agents should possess interactive capabilities to align with human intentions through interaction with users.
For these two different settings, in ColorAgent, we explore two plug-and-play modules~\citep{wu2025quick,wu2025verios}: personalized user intent recognition and proactive engagement.\footnote{The implementation details are available at \url{https://github.com/MadeAgents/Quick-on-the-Uptake} and \url{https://github.com/Wuzheng02/VeriOS}.}

\subsection{Personalized User Intent Recognition}
In the setting with additional user memory, we explore a recognizer framework~\citep{wu2025quick} that can rewrite user-provided queries into user-specific personalized queries and generate personalized standard operating procedures (SOPs) based on the additional user memory. 
The framework consists of two phases: the intention flow extraction phase and the deployment phase.

In the intention flow extraction phase, the framework retrieves users’ historical trajectories and extracts the corresponding SOPs for each query from these trajectories to construct a query-level explicit user intent knowledge base, while simultaneously refining user profiles through the analysis of historical trajectories to build a user-level implicit user intent knowledge base.

In the deployment phase, the framework first matches the current user query with the most similar query and SOP from the explicit user intent knowledge base using a retrieval-augmented generation (RAG)-based approach and extracts the SOP for the current user query via an SOP Extractor.
It then rewrites the current user query and SOP into a personalized user query and SOP using a query rewriter.
Therefore, the framework can incorporate relevant knowledge from both the explicit and implicit user intent knowledge bases into the operational mobile agent through personalized queries and SOPs, thereby achieving better alignment with human intent.

\subsection{Proactive Engagement}
In the setting without additional user memory, we explore a query-driven proactive human-agent-GUI interaction method~\citep{wu2025verios} to align with human intention by proactive engagement. 
And we employ a two-stage learning paradigm that facilitates the decoupling and utilization of meta-knowledge to construct the ask agent.

We categorize the required capabilities of the ask agent into two types of meta-knowledge: 
(i) the judgment of when to ask questions in a scenario and how to pose them, and (ii) the enhancement of action generation based on question-answer pairs to align with human intentions.
In our training framework, we decouple the same screenshot based on meta-knowledge into two samples. 
Specifically, the prompt for the first sample does not include the history of question-answer pairs, and if it is determined that the current scenario does not require asking a question, it will output a non-ASK action normally; 
if the current scenario is deemed to require a question, the action will be changed to an ASK action. 
The prompt for the second sample, however, includes the history of question-answer pairs, and the action is always a non-ASK action. 
We then conduct interleaved training with these two types of samples to obtain the final ask agent that incorporates both types of knowledge.

The ask agent is specifically designed for human-agent-GUI interaction and can autonomously determine whether the current scenario is trustworthy.
If the current scenario is trustworthy, the agent executes tasks automatically; if not, it aligns with human intent by proactively asking users for clarification.
Therefore, we dynamically bridge the gap between full automation and precise human intent alignment by intelligently deciding when to execute tasks and when to seek guidance.

\subsection{Experiment}
\begin{table}[htbp]
\centering
\begin{tabular}{lcc}
\toprule
\textbf{Methods} & \textbf{MobileIAR (IAR)} & \textbf{VeriOS-Bench (SR)} \\
\midrule
Qwen2.5-VL-3B-Instruct &12.07 & 5.35 \\
OS-Atlas-Pro-7B & 36.11 & 44.39\\
UI-TARS-7B-SFT & 37.05 & 49.73 \\
UI-TARS-7B-DPO & 36.19 & 49.73 \\
UI-TARS-1.5-7B & 36.88 & 48.13\\
Qwen2.5-VL-7B-Instruct & 15.30 & 23.53  \\
Qwen2.5-VL-32B-Instruct &  37.32 & 45.45  \\
Qwen2.5-VL-72B-Instruct & 53.75 &  54.01 \\
Qwen-VL-max & 17.99 & 18.72 \\
GPT-4o & 31.57 & 40.64 \\
\midrule
\textbf{Ours} & \textbf{58.66} & \textbf{68.98}\\
\bottomrule
\end{tabular}
\caption{Experimental results on MobileIAR and VeriOS-Bench. Our explored methods achieve optimal performance.}
\label{tab:tool2partner}
\end{table}

To validate the practical value of our explored methods, we conduct experiments on MobileIAR~\citep{wu2025quick} (with addtional user memory setting) and VeriOS-Bench~\citep{wu2025verios} (without addtional user memory setting).
MobileIAR is a user-specific OS agent benchmark with different ground truth annotations for different users, designed to test the capabilities of personalized OS agents. 
VeriOS-Bench, on the other hand, is an OS agent benchmark that includes a large number of untrustworthy scenarios, reflecting the trustworthiness of OS agents.  

We conduct tests on these two benchmarks, with MobileIAR reporting IAR (intention alignment rate) and VeriOS-Bench reporting SR (step-wise success rate).
Here, IAR requires the agent's output action to match the most intention-aligned action annotated in the MobileIAR dataset for the current user, while SR requires the output action to align with the ground truth action annotated in VeriOS-Bench.

As shown in Table~\ref{tab:tool2partner}, the experimental results show that our methods outperform popular OS agents, regardless of the availability of additional user memory. 
This represents a key step in transforming OS agents from mere tools into warm partners.

\section{Future Work}

Our ColorAgent represents an initial step toward building an OS Agent that can seamlessly interact with both the environment and the user. While our work demonstrates promising progress, constructing an agent that is stable, reliable, and entirely trustworthy in real-world scenarios remains an ambitious challenge. To ultimately evolve into a super-intelligent AI assistant that users can depend on for long-term interaction, several critical issues must be addressed. 

\textbf{Evaluation Paradigm.} Although our ColorAgent achieves impressive results on existing benchmarks, it's important to note that current benchmarks remain inadequate for a comprehensive evaluation of OS Agents. (i) There exists a substantial disparity between benchmark tasks and the complexities of real-world scenarios. Current benchmarks involve a limited range of applications, are dominated by simple tasks, and fail to reflect the user demands of complex tasks. Furthermore, they barely consider exceptional or unpredictable situations, which are intrinsic to real environments. (ii) Current evaluation focuses narrowly on task success rates, while neglecting user-centered dimensions such as the accuracy of intent recognition, the capacity for self-evolution through sustained interaction, and the overall quality of user experience. 
To mitigate these deficiencies, we are developing a novel benchmark designed to more faithfully approximate real-world scenarios, which will provide a more reliable evaluation environment and guide the development of practical systems for OS Agents.

\textbf{Agent Collaboration.} While our proposed multi-agent framework has demonstrated clear advantages over single-agent approaches, the design space for multi-agent collaboration remains largely unexplored. Future work includes exploring different collaborative architectures, such as centralized, sequential, or fully connected multi-agent systems~\citep{yang2024llm}, each offering different trade-offs in terms of scalability, flexibility, and communication overhead. Moreover, the issue of collaboration penalties among agents deserves further investigation. While cooperation can improve overall task coverage, it may also constrain the autonomy of individual agents and introduce efficiency bottlenecks. In the future, we aim to develop a more efficient multi-agent collaboration paradigm, offering scalable and robust solutions for OS Agents.

\textbf{Security.} In ColorAgent, we have introduced preliminary security mechanisms by enabling the agent to proactively query humans in untrustworthy scenarios. However, the large-scale deployment of OS Agents in real-world environments demands more comprehensive safeguards. One key direction is designing a safe and controllable sandbox environment, which allows the agent to satisfy user requirements while preventing unintended damage under abnormal conditions. Moreover, the ability of OS Agents to handle exceptional scenarios should be further strengthened to improve robustness during execution. Finally, fine-grained permission control is needed to clearly define the capability and operational boundaries of OS Agents~\citep{wu2025gem}, thereby avoiding excessive intervention in both the environment and the user’s activities.

\section{Conclusion}

In this report, we introduce ColorAgent, a mobile OS agent that supports both long-horizon, robust environment interaction and personalized, proactive user interaction. By combining model training with step-wise reinforcement learning and self-evolving mechanisms, together with a carefully designed multi-agent framework, ColorAgent achieves long-horizon, robust interactions with dynamic environments. Furthermore, our exploration of user intent recognition and human-agent interaction enables ColorAgent to transcend the role of a mere task-execution tool, evolving toward a warm and human-aligned OS agent. 

\section*{Acknowledgments}
\label{sec:ack}
We acknowledge the support of the SJTU-OPPO Joint Lab on Artificial Intelligence.


\bibliography{iclr2025_conference}
\bibliographystyle{rlc}

\clearpage
\appendix

\section{Action Space}\label{app:action-space}

\begin{table}[h]
\centering
\begin{tabular}{p{2.5cm}p{2.5cm}p{7.5cm}}
\midrule
\textbf{Action Type} & \textbf{Parameters} & \textbf{Description} \\
\midrule
\texttt{click} & \texttt{coordinate} & Click the screen at the specified \texttt{(x, y)} coordinate. \\
\midrule
\texttt{long\_press} & \texttt{coordinate}, \texttt{time} & Press and hold on the screen at \texttt{(x, y)} for a specified number of seconds. \\
\midrule
\texttt{swipe} & \texttt{coordinate}, \texttt{coordinate2} & Swipe from the starting coordinate \texttt{(x, y)} to the end coordinate \texttt{(x2, y2)}. \\
\midrule
\texttt{type} & \texttt{text} & Input text into the currently focused input box. \\
\midrule
\texttt{clear\_text} & \textbackslash & Clear the content of the active input box. \\
\midrule
\texttt{system\_button} & \texttt{button} & Press a system button: \texttt{Back}, \texttt{Home}, \texttt{Menu}, or \texttt{Enter}. \\
\midrule
\texttt{open} & \texttt{text} & Launch an app on the device by name. \\
\midrule
\texttt{wait} & \texttt{time} & Wait for a specified number of seconds. \\
\midrule
\texttt{answer} & \texttt{text} & Answer the user query. \\
\midrule
\texttt{terminate} & \texttt{status} & Terminate the current task and report whether it was a \texttt{success} or \texttt{failure}. \\
\midrule
\end{tabular}

\caption{Action space.}
\label{tab:action-space}

\end{table}

\section{Details of Each Discriminator}\label{app:discriminator}
The trajectory filtering module employs multiple specialized discriminators, each evaluating a specific dimension of trajectory quality. The discrimination aspects are:
\begin{itemize}[leftmargin=15pt]  
    \item \textbf{Task Completion}: Assesses whether the trajectory successfully achieves the targeted task by comparing the final GUI state with the expected outcome detailed in the query description.

    \item \textbf{Action Validity}: Evaluates the accuracy of individual actions, including the validity of coordinates for click actions, semantic alignment for text inputs, and adherence to established GUI operational rules.

    \item \textbf{Path Relevance}: Determines the logical necessity of each step in the trajectory, ensuring that actions are relevant and contribute meaningfully to task progress.

    \item \textbf{Reasoning Coherence}: Checks for logical consistency in consecutive actions, confirming that preceding steps effectively support the transition into subsequent operations.

    \item \textbf{Redundancy Check}: Identifies and reduces excessive actions within the trajectory by evaluating the necessity of each operational step, thereby minimizing redundant operations and repetition of erroneous actions.

    \item \textbf{User-Centric Evaluation}: Assesses task success from the user's perspective, prioritizing user experience and satisfaction with the achieved results.

    \item \textbf{Behavioral Analysis}: Evaluates the agent's goal-directed behavior and decision-making quality, assessing adaptability to changing conditions.
\end{itemize}

\section{Benchmark Issue Fixes}
\label{app:fix-benchmark}
\textbf{AndroidWorld} When conducting multiple experiments using the same Android Virtual Device (AndroidWorldAVD), we observed that the internal state of certain apps could vary depending on tasks executed in previous runs (for example, the camera might remain in photo or video mode). Although a robust agent should ideally handle all possible states, such variations can introduce instability and inconsistency in evaluation. To mitigate this issue, we implemented a modification: before executing any task involving \texttt{Audio Recorder}, \texttt{Camera}, \texttt{Tasks}, \texttt{Markor}, \texttt{Simple Calendar Pro}, or \texttt{Chrome} apps, we reset the corresponding app to ensure a consistent internal state across all task runs.

\textbf{AndroidLab} When performing tasks related to \texttt{Clock}, we fixed an issue that prevented the correct extraction of information from the clock interface. For \texttt{Settings}-related tasks, we addressed problems that caused failures in retrieving the app storage, system brightness, and app notification information.

Our evaluation code on the AndroidWorld and AndroidLab benchmarks is available at \url{https://github.com/MadeAgents/mobile-use}.

\end{document}